\documentclass[fleqn,twoside]{article}
\usepackage{espcrc2}
\usepackage{graphicx}

\usepackage{graphicx}
\usepackage[figuresright]{rotating}

\newcommand{\AmS}{{\protect\the\textfont2
  A\kern-.1667em\lower.5ex\hbox{M}\kern-.125emS}}

\title{Parameter degeneracies and
new plots in neutrino oscillations
}

\author{Osamu Yasuda\\
        {\ }\\
        Department of Physics, Tokyo Metropolitan University\\
        1-1 Minami-Osawa Hachioji, Tokyo 192-0397, Japan}
       
\begin{document}
\pagestyle{empty}

\begin{abstract}
It is shown that plots of constant probabilities in the
$(\sin^22\theta_{13}, 1/s^2_{23})$ plane enable us to
see eightfold degeneracy easily.
Using this plot,
I discuss how an additional long baseline
measurement resolves degeneracies after the JPARC
experiment measures the oscillation probabilities
$P(\nu_\mu\rightarrow\nu_e)$ and
$P(\bar{\nu}_\mu\rightarrow\bar{\nu}_e)$ at $|\Delta m^2_{31}|L/4E=\pi/2$.

\vspace{1pc}
\end{abstract}

\maketitle

\begin{figure}
\vglue -0.5cm
\hglue -.5cm
\includegraphics[scale=0.46]{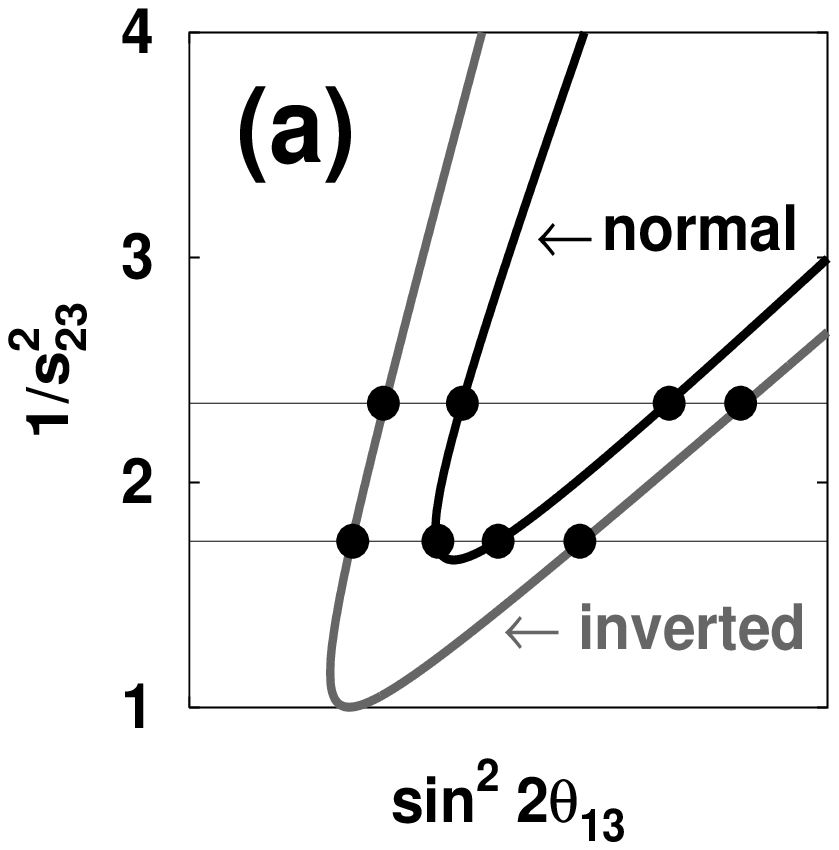}
\vglue -4.7cm
\hglue 3.5cm
\includegraphics[scale=0.5]{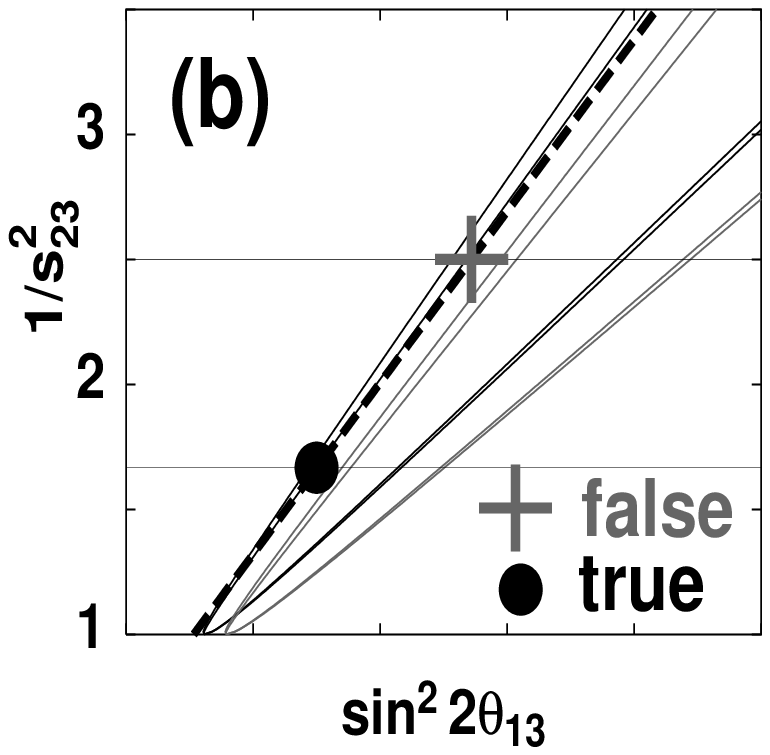}
\vglue -1.cm
\caption{(a) Trajectories of solutions given by
$P(\nu_\mu\rightarrow\nu_e)=$const. and
$P(\bar{\nu}_\mu\rightarrow\bar{\nu}_e)=$const.,
and the eightfold degeneracy.
(b) Trajectories given by the third experiment
on $\nu_\mu\rightarrow\nu_e$
(solid lines) and the JPARC experiment (dashed
line), where black (gray) lines are for normal
(inverted) hierarchy, and the blob (cross)
stands for the true (false) solution.}
\label{fig}
\end{figure}

It is known that even if the values of the oscillation
probabilities $P(\nu_\mu \rightarrow \nu_e)$ and
$P(\bar{\nu}_\mu \rightarrow \bar{\nu}_e)$ are exactly given,
we cannot determine uniquely the values of the
oscillation parameters due to eightfold parameter degeneracy.
To see how the eightfold degeneracy is lifted, it is necessary for
the plot to give eight different points for different eight solutions.
In Ref. \cite{Yasuda:2004gu} it was shown that the solution
given by $P(\nu_\mu \rightarrow \nu_e)=$const. and
$P(\bar{\nu}_\mu \rightarrow \bar{\nu}_e)=$const. gives
a hyperbola in the $(\sin^22\theta_{13}, 1/s^2_{23})$ plane
in most cases,
as is shown in Fig.\ref{fig}(a), so that this plot is useful
to see how the eightfold degeneracy is resolved.

It is expected that the JPARC
experiment will measure the oscillation probabilities
$P(\nu_\mu\rightarrow\nu_e)$ and
$P(\bar{\nu}_\mu\rightarrow\bar{\nu}_e)$ at $|\Delta m^2_{31}|L/4E=\pi/2$.
Here I would like to discuss how the eightfold degeneracy is resolved
by an additional experiment.
From the JPARC experiment on $\nu_\mu\rightarrow\nu_e$ and
$\bar{\nu}_\mu\rightarrow\bar{\nu}_e$ at $|\Delta m^2_{31}|L/4E=\pi/2$,
we can deduce the value of the CP phase $\delta$
up to the eightfold ambiguity.  The trajectories given by the JPARC experiment
at $|\Delta m^2_{31}|L/4E=\pi/2$ turn out to be straight lines.
For each value of $\delta$,
from the third experiment on $\nu_\mu\rightarrow\nu_e$
(or $\bar{\nu}_\mu\rightarrow\bar{\nu}_e$)
one obtains a unique trajectory in the
$(\sin^22\theta_{13}, 1/s^2_{23})$ plane.
Putting the trajectories of the third and JPARC experiments together
in the $(\sin^22\theta_{13}, 1/s^2_{23})$ plane,
we find that (i) in general it is difficult to resolve
the $\theta_{23}$ ambiguity
as is shown in Fig.\ref{fig}(b)
(this is not the case with the silver channel $\nu_e\rightarrow\nu_\tau$),
(ii) the third experiment
with longer baseline and lower energy resolves the
sgn($\Delta m^2_{32}$) ambiguity better, and (iii) the $\delta\leftrightarrow
\pi-\delta$ ambiguity is resolved better for
$\pi/2<|\Delta m^2_{31}|L/4E<\pi$.

This work was supported in part by Grants-in-Aid for Scientific Research
No.\ 16540260 and No.\ 16340078, Japan Ministry
of Education, Culture, Sports, Science, and Technology.

\begin{thebibliography}{9}
\bibitem{Yasuda:2004gu}
O.~Yasuda,
New J.\ Phys.\  {\bf 6} (2004) 83
[arXiv:hep-ph/0405005].
\end{thebibliography}
\end{document}